\documentclass[a4paper,100pt]{article}
\usepackage{graphicx} 
\usepackage{subfigure}
\usepackage[utf8]{inputenc}
\title{Lagrangian Formulation of the Snowplow Model and Operating Point for Z pinch Devices}
\author{Miguel Cárdenas, Alejandro Nettle and Leandro Nú\~{n}ez\\Universidad de Playa Ancha\\Av. Playa Ancha 850, Valparaíso, Chile\\e-mail: miguel.cardenas@upla.cl}

\begin{document}
\date{}

\large
\maketitle
\begin{abstract}
We write down the lagrangian for the snowplow model of the Z pinch system. Then, we develop the Euler-Lagrange equations to find out the corresponding equations of motion. Next, we set a criterion for quantitatively estimating the performance of Z pinch devices. Finally, we apply this criterion to a specific Z pinch system.
\end{abstract} 

\section{Introduction}

Since the beginning of controlled thermonuclear fusion research, magnetic confinement has been a key component \cite{bishop}\cite{hagler}\cite{post}\cite{kolb}. This method is based on the pinch effect a process where the self-magnetic field of a current-carrying plasma between two electrodes compresses the plasma, heating and confining it \cite{bennett}\cite{tonks}\cite{jackson}\cite{reitz-milford}\cite{krall}. In theory, heating and confining the plasma by using the pinch effect would require no auxiliary magnetic field. However, in practice a current-carrying plasma often becomes unstable due to  magnetohydrodynamic effects, a phenomenon that requires the use of additional magnetic fields to maintain stability \cite{glasstone}. Another problem is that heating and confining the plasma at the same time can become conflicting goals. In effect, there are two mechanisms for heating the plasma, namely ohmic
heating and adiabatic compression heating. A poorly ionized plasma exhibits a paradoxical relationship where the low degree of ionization increases ohmic heating but decreases plasma conductivity, thereby weakening its magnetic field. Therefore, this weak magnetic field, which only influences charged particles, is less effective at both containing the plasma and heating it adiabatically.

In this work, we will be concerning with the Z pinch configuration which has been regarded and it still is as a promising one for the goal of achieving controlled thermonuclear fusion \cite{ryutov}. Our approach to the subject is via mathematical modeling of that system. In particular, we adopt the snowplow model where the working plasma is assumed to be fully ionized, making its conductivity infinite. The model simplifies the current-carrying plasma into an infinitely thin cylindrical layer, which acts as a piston that pushes the plasma through the discharge. So, in this model the possibility of the onset of instabilities, \textit{e.g.} sausage instability, kink instability and so on, during discharges is excluded \cite{rosenbluth}. 

The snowplow model has been extensively used over the decades for studying the electrical and dynamical items of the discharges in Z pinch devices. 
Comparison of those results with data gathered from real parallel experiments is rather satisfactory. Recently, we adapted the snowplow model to also obtain information about the thermal behavior of the discharge \cite{cardenas}. Nonetheless, a comparative study of the thermal behavior of discharges as predicted by the model versus that obtained from experiments is still pending.

Technically speaking, the snowplow equations, vis-\`a-vis the modified snowplow equations, consist of a set of two coupled  nonlinear integro-differential equations for the radius of the current sheath and the plasma current through it as a function of time. Although these equations are very complicated, they can be obtained easily by using elementary notions like Newton's second law and basic knowledge on electrical circuits \cite{rosenbluth}\cite{cardenas}. Despite this, we think it would be worthwhile to try to take a more comprehensive look at this problem. 

Another aspect we are interested in investigating concerns the possible connection between the fed energy of the system and its performance. Thus, in this article, we focus on these two subjects. First, we look for the possibilty of placing the snowplow equations in a more general context. We address this issue by appealing to the lagrangian formulation of the problem.  At the first place, we set the suitable lagrangian for the Z pinch system. Then, we apply the Euler-Lagrange equations to that lagrangian. As a results, we obtain the modified snowplow equations and also as a particular case the original snowplow equations. So, now we have as a byproduct at disposal all the related theoretical machinery, \textit{e.g.} symmetry, invariance, extension of the Noether's theorem and so on, to get new results and a better insight on the subject. 

To approach the other topic of our interest, we devise three quantifiers that on the basis of the data gathered from numerical simulations can provide undeniable information about the performance of Z pinch devices. Then, we use these model-based tools for analyzing the performance of a specific Z pinch setup. What we do is sweeping the charging voltage of the chosen Z pinch system while keeping the rest of its parameters fixed. In this manner, we discovered that for each Z pinch setup there exists a unique source-energy value that optimizes the performance of that apparatus. We refer to this source-energy value as the operating point of the system because it ensures the best source-load coupling. 

The organization of the paper is the following: Section 2 and Section 3 are devoted to set and to develop the lagrangian formulation for the Z pinch system. Section 4 concerns the general theory and methods for analyzing the performance of Z pinch devices. Then, an application of those methods to a specific case is fully reported. In Section 5, we summarize the main conclusions reached in this research work.

\section{Lagrangian formulation of the snowplow model}

Within the framework of the snowplow model, the dynamics of the Z pinch discharges can be appropriately accounted for by means of just two dynamical variables. Such variables correspond to the radius of the plasma sheath $r(t)$ and the charge stored in the bank of capacitors $Q(t)$  where $t$ stands for time. 

To construct the lagrangian for this system, we have first to set clearly, in terms of $r$, $\dot r\equiv(dr/dt)$, $Q$ and $\dot Q\equiv(dQ/dt)$, what the kinetic and potential energy of the system are. In this work, we adopt the choice of defining the kinetic energy as

\begin{equation}
\mathcal T(r,\dot r,\dot Q)=\frac{1}{2}M(t)\,\dot r^2+\frac{1}{2}L(t)\,\dot Q^2
\end{equation}
where

\begin{equation}
M(t)=\rho_0\, l_0\,\pi(r_0^2-r^2)
\end{equation}
corresponds to the mass of the current sheath whereas 

\begin{equation}
L(t)=L_0-\left(\frac{\mu_0l_0}{2\pi}\right)\ln{\left(\frac{r}{r_0}\right)}
\end{equation}
represents the total inductance of the system. Here, $\rho_0$ accounts for the filling density of the cylindrical vessel, $r_0$ is the inside radius of the vessel, $l_0$ stands for the length of the cylinder and $L_0$ corresponds to the parasitic inductance of the setup.

What motivates our particular choice for kinetic energy is that in that way potential energy becomes simpler. In effect, in this manner the potential energy only includes the term that accounts for the electrostatic energy stored in the bank of capacitors but  does not include the magnetic energy stored in the inductance \cite{lindsay}. In other words, potential energy depends only on the generalized coordinate $Q$ but does not depend on the generalized velocity $\dot Q$. The specific expression for the potential energy is

\begin{equation}
\mathcal V(Q)=\frac{1}{2C_0}Q^2
\end{equation}
where $C_0$ represents the capacity of the bank of capacitors. So, the lagrangian for the Z pinch system is given by the expression

\begin{equation}
\mathcal L=\frac{1}{2}M\,\dot r^2+\frac{1}{2}L\,\dot Q^2-\frac{1}{2C_0}Q^2.
\end{equation}

This dynamical system displays two distinctive characteristics that require a dedicated treatment. One of these is that the mass of the current sheath $M$ depends explicitly on the dynamical variable $r$, so that $M$ is not constant but varies as the discharge progresses. The other salient feature is that the Z pinch dynamical system as a whole is not conservative. Naturally, it is worth highlighting that total energy, in its broadest thermodynamic sense, is conserved. In effect, the piece of organized energy that is irretrievably deconstructed or ''lost'' during the progress of the discharge, however, re-emerges in the form of internal energy of the plasma that from the operational point of view obeys the formula \cite{cardenas}

\begin{equation}
U(t)=-\pi r_0^2 l_0\rho_0\int_0^t\frac{1}{r}\left(\frac{dr}{dt'}\right)^3 dt'.
\end{equation}
This internal energy entails in turn kinetic pressure in the plasma which manifests as a force, $F_k(t)$ let us say, pushing the current sheath outwards. Thus, the magnetic force, \textit{i.e.} the $J\times B$ force, that pushes the current sheath inwards in combination with the force $F_k(t)$ is what finally determines the dynamics of the current sheath. The explicit expression for the generalized force $F_k(t)$ is

\begin{equation}
F_k(t)=-\frac{4}{3}\rho_0\, l_0\,\pi\, r_0^2\,\left(\frac{1}{r}\right)\int_0^t \left(\frac{\dot r^3}{r}\right) dt'.
\end{equation}

\section{The Euler-Lagrange equations}
Our dynamical system of variable mass $M$ subject to the generalized force $F_k$ is well described by means of the following set of Euler-Lagrange equations \cite{corben}\cite{pesce}:

\begin{equation}
\frac{d}{dt}\left(\frac{d\mathcal L}{d\dot r}\right)-\left(\frac{d\mathcal L}{dr}\right)=F_k(t)-\frac{1}{2}\left(\frac{dM}{dr}\right)\,\dot r^2
\end{equation}
and

\begin{equation}
\frac{d}{dt}\left(\frac{d\mathcal L}{d\dot Q}\right)-\left(\frac{d\mathcal L}{dQ}\right)=0.
\end{equation}

Now, keeping in mind that the relationship between the charge in the capacitor bank, $Q(t)$, and the current that flows through the current sheath, $I(t)$, is given by 

\begin{equation}
Q(t)=Q_0-\int_0^t I(t')\,dt',
\end{equation}
we obtain the relation $\dot Q(t)=-I(t)$. Here, $Q_0$ stands for the charge initially stored in the bank of capacitors which is determined by the charging voltage of the capacitor bank $V_0$ as $Q_0=C_0V_0$.

The development of the Euler-Lagrange equations -\textit{i.e.} Eqn.(8) and Eqn.(9)- and further conversion of them to dimensionless variables yields

\begin{equation}
\left(\frac{d^2r}{dt^2}\right)=\frac{6r^2 (dr/dt)^2-3\alpha^2 I^2-4\int_0^t (1/r)(dr/dt')^3 dt'}{3r(1-r^2)}
\end{equation}
and

\begin{equation}
\left(\frac{dI}{dt}\right)=\frac{1-\int_0^tIdt'+\beta (I/r)(dr/dt)}{1-\beta\ln{(r)}}
\end{equation}
where $t$, $r$ and $I$ are now dimensionless variables obtained from dividing the physical time by $\sqrt{L_0C_0}$, the radius of the current sheath by $r_0$ and the current flowing along the current sheath by $V_0\,\sqrt{C_0/L_0}$, respectively. The dimensionless parameters $\alpha$ and $\beta$ are given in terms of the physical specifications of the corresponding experiment by

\begin{equation}
\alpha=\sqrt{\frac{\mu_0 C_0^2  V_0^2}{4\pi^2  r_0^4\rho_0 }}
\end{equation}
and 
\begin{equation}
\beta=\left(\frac{\mu_0 l_0}{2\pi L_0}\right).
\end{equation}

To actually solve the equations displayed above, we have to supplement them with the following initial conditions:
\begin{equation}
r(0)=1,
\end{equation}

\begin{equation}
I(0)=0,
\end{equation}

\begin{equation}
\left(\frac{dr}{dt}\right)_{t=0}=0,
\end{equation}

\begin{equation}
\left(\frac{d^2r}{dt^2}\right)_{t=0}=-\left(\frac{\alpha}{\sqrt{3}}\right)
\end{equation}
and

\begin{equation}
\left(\frac{dI}{dt}\right)_{t=0}=1.
\end{equation}

As we might expect, Eqn.(11) and Eqn.(12) correspond to the modified snowplow equations we have obtained elsewhere by a different procedure \cite{cardenas}. Moreover, when we ignore at all the kinetic pressure -\textit{i.e.} when we put $F_k=0$ on the right hand side of Eqn.(8)-  the Euler-Lagrange equations lead to

\begin{equation}
\left(\frac{d^2r}{dt^2}\right)=\frac{2r^2 (dr/dt)^2-\alpha^2 I^2}{r(1-r^2)}
\end{equation}
and

\begin{equation}
\left(\frac{dI}{dt}\right)=\frac{1-\int_0^tIdt'+\beta (I/r)(dr/dt)}{1-\beta\ln{(r)}}
\end{equation}
which exactly match the equations derived within the original formulation of the snowplow model \cite{rosenbluth}\cite{cardenas}. 
 
\section{Analysis of the performance of Z pinch devices}

In this section, we use the results obtained from the snowplow scheme to analyze energy exchanges in discharges through Z pinch devices. Essentially, we analyze quantitatively how the capacitor bank transfers energy during discharges in Z pinch devices.
Next, we present with great deal of detail a specific application that allows us to discover the optimal conditions for heating adiabatically and containing a plasma during a discharge in a Z pinch device.

\subsection{Diagnostic tools}

To start with, we quote that the energy delivered by the bank of capacitors, over the period of time $0\to t$, to its attached Z pinch device is given by the expression \cite{cardenas2}

\begin{equation}
E_d(t)=2E_0\left(-\beta\int_0^t\frac{1}{r}\left(\frac{dr}{dt'}\right)I^2dt'
+\int_0^t\left(1-\beta\ln{(r)}\right)I\left(\frac{dI}{dt'}\right)dt'\right)
\end{equation}
where $E_0$ stands for the energy initially stored in the bank of capacitors which is computed as

\begin{equation}
E_0=\frac{1}{2}C_0V_0^2.
\end{equation}
Now, a portion of the energy $E_d$ gets stored in the inductor of inductance $L$ while the rest $E_i$ is what is actually transferred to the plasma as mechanical energy. That part is given by the expression \cite{cardenas2}

\begin{equation}
E_i(t)=-E_0\,\beta\int_0^t\frac{1}{r}\left(\frac{dr}{dt'}\right)I^2dt'.
\end{equation}
In turn, not all this energy converts to internal energy of the plasma but only a part of it, namely $U(t)$, is what really becomes internal energy of the plasma. That piece of $E_i(t)$ was already quoted through Eqn.(6) and it can also be expressed, for convenience, in terms of dimensionless $t$ and $r$ as \cite{cardenas2}

\begin{equation}
U(t)=-E_0\,\left(\frac{\beta}{\alpha^2}\right)\int_0^t\frac{1}{r}\left(\frac{dr}{dt'}\right)^3 dt'.
\end{equation}

Thus, in general terms, the performance or efficiency of a given configuration can be quantified by means of the quotient

\begin{equation}
\eta_1=\left(\frac{E_d}{E_0}\right)\times 100
\end{equation}
what becomes

\begin{equation}
\eta_1=2\left(-\beta\int_0^t\frac{1}{r}\left(\frac{dr}{dt'}\right)I^2dt'
+\int_0^t\left(1-\beta\ln{(r)}\right)I\left(\frac{dI}{dt'}\right)dt'\right)\times 100
\end{equation}
according to Eqn.(22) and implicitly depends on $E_0$.

This quantity tells us what percentage of the energy initially stored in the bank of capacitors, $E_0$,  is actually transferred to its attached Z pinch device during the period $0\to t$. Before proceeding, we must point out that the function $\eta_1$ depends on two parameters, namely
the value of the energy initially stored in the capacitor bank $E_0$ and the extension of the integration interval $0\to t$ considered in Eqn.(27). In the simulations that follow, we will compute $\eta_1$ for different values of $E_0$. The integration interval $0\to t$ considered will be specific to each case and we will choose it in a way that $t=t_p$. In other words, $t$ will match always the time when the first pinch of the plasma sheath happens for each value $E_0$. 

In an analogous manner, we introduce two new efficiency quantifiers, namely 

\begin{equation}
\eta_2=\left(\frac{E_i}{E_0}\right)\times 100
\end{equation}
what becomes 

\begin{equation}
\eta_2=-\left(\,\beta\int_0^t\frac{1}{r}\left(\frac{dr}{dt'}\right)I^2dt'\right)\times 100
\end{equation}
according to Eqn.(24) and implicitly depends on $E_0$ and 

\begin{equation}
\eta_3=\left(\frac{U}{E_0}\right)\times 100
\end{equation}
that becomes

\begin{equation}
\eta_3=-\left(\,\left(\frac{\beta}{\alpha^2}\right)\int_0^t\frac{1}{r}\left(\frac{dr}{dt'}\right)^3 dt'\right)\times 100
\end{equation}
according to Eqn.(25) and also implicitly depends on $E_0$.

These quantifiers are obvious; $\eta_2$ accounts for the percentage of $E_0$ that enters the plasma during the period $0\to t$ whereas $\eta_3$ measures the percentage of $E_0$ that really transforms to internal energy of the plasma during the period $0 \to t$. Let us emphasize once again that the evaluation of the integrals in Eqn.(27), Eqn.(29) and Eqn.(31) is carried out by setting
$t=t_p$ as the upper limit of each of those integrals.

\subsection{Applying the method to a specific situation}

We consider a prototype Z pinch device with a working voltage of the order of few kilovolts. More specifically, the cylindrical vessel of the prototype is a fraction of a meter long and of about $3\,L$ volume. This vessel is then filled with helium gas at $1\, mm$ of mercury pressure approximately. We consider that the parasitic inductance of the setup is $L_0\approx 30\,[nH]$ and the total capacity of the capacitor bank is $C_0\approx\,80 [\mu F]$.

Our target is to study the behavior of discharges in this device along a wide range of fed energy. To achieve it in a real experiment, the experimenter has only to sweep the charging voltage at the capacitor bank over a wide range. That procedure is exactly what we do in our parallel numerical simulations. Thus, for each voltage $V_0$ we compute $E_0$ by using Eqn.(23). After that, we  solve numerically Eqn.(11) and Eqn.(12) to look for the time $t_p$ where the curve $r(t)$ reaches its first minimum.
With that information at hand, we determine according with Eqn.(27), Eqn.(29) and Eqn.(31) the value of $\eta_1, \eta_2$ and $\eta_3$ for each particular value of the source energy $E_0$, respectively. The corresponding curves are displayed in the Figure 1.

The curve $\eta_1$ \textit{vs} $E_0$ shows an unmistakable maximum. This maximum occurs at $E_0\approx 0.33\,[kJ]$ vis-\`a-vis $V_0\approx 2.8\,[kV]$ and its height equals $100\,\%$. The corresponding temperature of the plasma is $k_BT\approx 5.12\,[eV]$. 

Thus, when $E_0\approx 0.33\,[kJ]$, the capacitor bank delivers all its electrostatic energy to the attached Z pinch device during the time interval $0\to t_p$. Hence, the configuration in which the source energy is tuned to the value $E_0\approx 0.33\,[kJ]$ is the one that leads to the best comprehensive coupling between source and load.
\begin{figure}[!h]
\centering
\includegraphics[scale=.8]{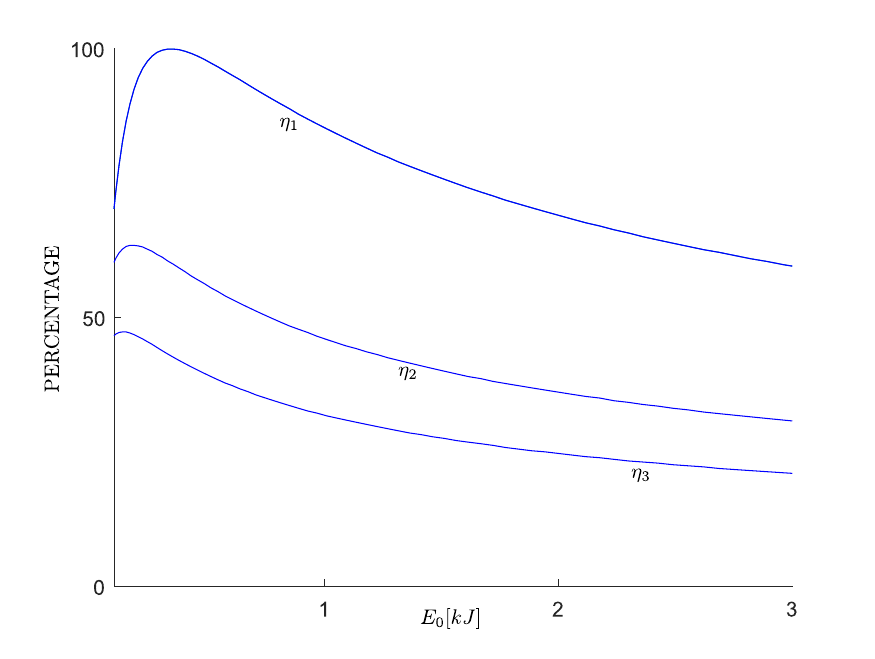}
\caption{The ratios $\eta_1$,$\eta_2$ and $\eta_3$ plotted against $E_0$.}
\end{figure}

On the other hand, curve $\eta_2$ \textit{vs} $E_0$ also exhibits a notable maximum but this is located at $E_0\approx 0.19\,[kJ]$  
vis-\`a-vis $V_0\approx 2.1\,[kV]$ and its height reaches the value $63\,\%$ approximately. The corresponding temperature of the plasma is $k_BT\approx\,3.12\,[eV]$. 

Therefore, the configuration where $E_0\approx 0.19\,[kJ]$ is the one that most effectively transforms the electrostatic energy of the capacitor bank into kinetic energy of the plasma.

However, in a view to revealing the conditions for a possible thermonuclear fusion process, the key information can be visualized more easily by examining curve $\eta_3$ \textit{vs} $E_0$. This curve also contains a maximum that is  located at $E_0\approx 0.14\,[kJ]$  vis-\`a-vis $V_0\approx 1.8\,[kV]$ and whose height is $47\,\%$ approximately. In this case, the corresponding temperature of the plasma is $k_BT\approx 2.32\,[eV]$.
Hence, the configuration where $E_0\approx 0.14\,[kJ]$ is the most effective in providing internal energy to the plasma at the expense of the electrostatic energy stored in the capacitor bank. We emphasize that the maxima of the three curves, $\eta_1$, $\eta_2$ and $\eta_3$ \textit{vs} $E_0$, are located relatively close to one another on the $E_0-$axis.

Incidentally, let us to point out that in the situation under examination where the number of particles in the container $N$ remains invariable, we can analogize internal energy and temperature because they are related through the formula

\begin{equation}
k_BT=\frac{2}{3}\left(\frac{U}{N}\right)
\end{equation}

Regarding the choice of the figure of merit associated with our simulations, we look at the curve $\eta_3$ because we want to discover the most efficient way to adiabatically heat the plasma. In this connection, we can assert that, according to our simulations, the optimal choice for the $E_0$ value of the system under scrutiny is given by the value of $E_0$ that peaks the $\eta_3$ \textit{vs} source-energy curve. Therefore, if we trust the simulations, we can venture that in the corresponding real experiment the best choice for the source-energy value should be $E_0\approx 0.14\,[kJ]$  vis-\`a-vis $V_0\approx1.8\,[kV]$.

At this point, it is relevant to point out that, according to Eqn.(30) and Eqn.(32), a configuration of initial stored energy $E_0$ and efficiency 
\begin{equation}
\eta_3=\left(\frac{U}{E_0}\right)\times 100
\end{equation}
leads to the following expression for the plasma temperature at pinching:

\begin{equation}
k_BT=\frac{2}{3}\left(\frac{E_0}{N}\right)\times \left(\frac{\eta_3}{100}\right).
\end{equation}
Therefore, since $\eta_3$ is never too small, we could state that the larger $E_0$ the larger $k_BT$. However, as far as real experiments are concerned, this statesment is not yet fully confirmed. In real experiments, when $E_0$ is very large, vis-\`a-vis when $V_0$ is high, there is no possibility of thermalization of the plasma because its collapse is too fast \cite{glasstone}. In contrast, our modeling of these experiments by construction always guarantees the thermalization of the plasma. This is clearly reflected in Eqn.(34).

In closing, let us recall that in a previous work, we postulated for large values of $E_0$ the following relationship between $E_0$ and the temperature of the plasma at pinching \cite{cardenas2}:

\begin{equation}
k_BT\propto E_0\,^{0.5}.
\end{equation}
Now that as a result of our simulations we know the curve $\eta_3$ \textit{vs} $E_0$ in full detail, we are in a position of perfecting that statesment. To do it, we fit the $\eta_3$ \textit{vs} $E_0$ curve at the region of interest. The result is 

\begin{equation}
\eta_3\approx 31\,E_0^{-0.36}
\end{equation}
which we show through Figure 2.

\begin{figure}[!h]
\centering
\includegraphics[scale=.8]{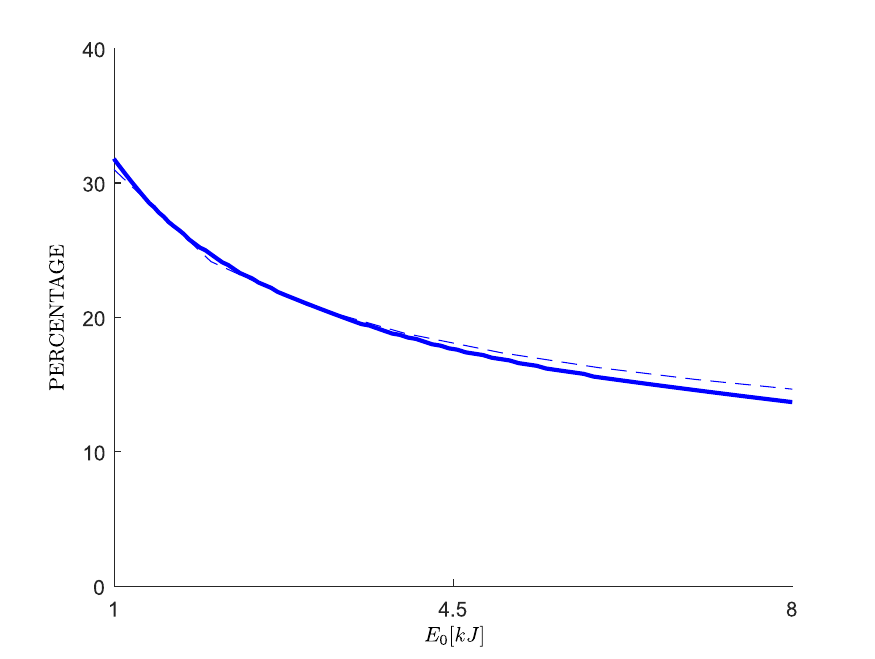}
\caption{The $\eta$ ratio plotted against $E_0$. Simulation: solid line, Fitting: dashed line.}
\end{figure}

After introducing this amendment into Eqn.(34), we find that for large values of $E_0$, the following relation holds:

\begin{equation}
k_BT\propto E_0\,^{0.64}.
\end{equation}
This relationship definitely fits better than Eqn.(35) the data gathered from our simulations. However, let us insist on this, whether or not Eqn.(37) is relevant can be judged only  by comparing it with its experimental counterpart.

Finally, let us reflect on the fact that throughout the present simulations as well as in previous simulations, the plasma temperature has never climbed to a substantial value. Of course, the exception to this rule occurs either when we exaggeratedly increase the source voltage or when we dramatically decrease the plasma density. However, in both those limit cases we confront an ultrafast pinch where the assumptions on which the snowplow model rests are no longer valid. 

At this point it is worth asking whether magnetic confinement using only the pinch effect, without external fields, is truly suitable for achieving high-temperature plasmas.
Moreover, unless the contrary is proven, it is not to be ruled out that thermonuclear fusion via the pinch effect could be just one more of the impossible things.

\section{Conclusions}

We list below what we believe are the most relevant conclusions from this research.
\begin{enumerate}
\item The dynamics of discharges in Z pinch devices can be formulated within the framework of the lagrangian theory. As proof of this, we make that lagrangian formulation available in this work.
\item We discovered that for each Z pinch configuration there is a single source-energy value that optimizes the performance of that specific Z pinch apparatus. We hypothesize that this result obtained in the snowplow-model context is extrapolable to real experiments. 
\item Of course, other pinch effect-based systems, like toroidal systems \cite{rosenbluth}, plasma focus systems \cite{mather}, etc, where the snowplow model applies should lead to the same result. 
\item We found that when the stored-energy value at the capacitor bank greatly exceeds its optimal value, the plasma temperature behaves as
\begin{equation}
k_BT\propto E_0\,^{0.64}.
\end{equation}
However, we believe this model-based relationship still requires confirmation based on real-world experiments.
\end{enumerate}


\begin{thebibliography}{99}
\bibitem{bishop}Amasa S. Bishop, 'Project Sherwood- The U. S. Program in Controlled Fusion', Addison-Wesley, Reading, Massachusetts (1958).
\bibitem{hagler}M. O. Hagler and M. Kristiansen, 'An Introduction to Controlled Thermonuclear Fusion', Lexington Books D. C. Heath and Company Lexington Massachusetts Toronto (1977).
\bibitem{post}Richard F. Post, Controlled Fusion Research- An Appplication of the Physics of High Temperature Plasmas, Rev. Mod. Phys. \textbf{28}, 338 (1956).
\bibitem{kolb}Alan C. Kolb, Magnetic Compression of Plasmas, Rev. Mod. Phys. \textbf{32}, 748 (1960).
\bibitem{bennett}W. H. Bennett: Magnetically Self-focusing Sreams, Phys. Rev. \textbf{45}, 890 (1934).
\bibitem{tonks}L. Tonks, 'Theory of Magnetic Effects in the Plasma of an Arc', Phys. Rev. \textbf{56}, 360 (1939). 
\bibitem{jackson}J. D. Jackson, 'Classical Electrodynamics', John Wiley and Sons, New York (1975).
\bibitem{reitz-milford}John R. Reitz and Frederick J. Milford, 'Foundations of Electromagnetic Theory', Addison-Wesley Publishing Company, Inc. Reading Massachusetts, U. S. A. (1960).
\bibitem{krall}N. A. Krall and A. W. Trivelpiece, 'Principles of Plasma Physics', McGraw-Hill, Inc., New York (1973).
\bibitem{glasstone}S. Glasstone and R. H. Lovberg: 'Controlled Thermonuclear Reactions'
, D. Van Nostrand Company, Princeton, N. J. (1960).
\bibitem{ryutov}D. D. Ryutov, M. S. Derzon and M. K. Matzen: The physics of fast Z pinches, Rev. Mod. Phys. \textbf{72}, 167 (2000).
\bibitem{rosenbluth}M. N. Rosenbluth, R. Garwin and A. Rosenbluth: Infinite Conductivity Theory of the Pinch, Report LA-1850, Los Alamos Scientific Laboratory, New Mexico, September 1954.
\bibitem{cardenas}Miguel Cárdenas, Alejandro Nettle and Leandro Núñez, Scaling Law for Discharges in Z pinch Devices, arXiv.org, 12 February 2025, [arXiv:2502.08570v1].
\bibitem{lindsay} Robert Bruce Lindsay, 'Concepts and Methods of Theoretical Physics', D. Van Nostrand Company, Inc., New York (1951).
\bibitem{corben} H. C. Corben and Philip Stehle, 'Classical Mechanics', John Wiley and Sons, Inc., New York (1960).
\bibitem{pesce}C. P. Pesce, The Application of Lagrange Equations to Mechanical Systems with Mass Explicitly Dependent on Position, J. Appl. Mech. \textbf{70}, 751 (2003).
\bibitem{cardenas2}Miguel Cárdenas, Alejandro Nettle and Leandro Núñez, Snowplow Model Predictions for Plasma Temperature in Z pinch Discharges, arXiv.org, 19 June 2025, [arXiv:2506.16551v1].
\bibitem{mather}J. Mather: An Intense Source of Neutrons from the Dense Plasma Focus, Intense Neutron Sources: Proceedings of a United States Atomic Energy Commission/European Nuclear Energy Agency seminar, Santa Fe, New Mexico, 19-23 September 1966.
\end{thebibliography}
\end{document}